\documentclass{article}

\usepackage{PRIMEarxiv}

\usepackage[utf8]{inputenc} 
\usepackage[T1]{fontenc}    
\usepackage{hyperref}       
\usepackage{url}            
\usepackage{booktabs}       
\usepackage{amsfonts}       
\usepackage{nicefrac}       
\usepackage{microtype}      
\usepackage{lipsum}
\usepackage{fancyhdr}       
\usepackage{graphicx}       
\graphicspath{{media/}}     

\title{Brain Tumor classification and Segmentation using Deep Learning
\thanks{\textit{\underline{Citation}}: 
\textbf{Romario , Dr Belal. Title. DOI:000000/11111.}} 
}

\author{
  Dr Belal Badawy , Romario Sameh Samir, Youssef Tarek \\
  Ahram Canadian University \\
   \And
   Mohammed Ahmed  , Rana Ibrahim , Manar Ahmed , Mohamed Hassan\\
  Ahram Canadian University \\
}

\begin{document}
\maketitle

\begin{abstract}
Brain tumor classification and segmentation is a project that involves using medical imaging techniques, such as magnetic resonance imaging (MRI) scans, to classify and segment different types of brain tumors. The goal of the project is to accurately identify and segment different types of brain tumors, such as gliomas and meningiomas, in order to improve diagnosis and treatment planning for patients with brain tumors. This is typically done using various machine learning and image processing techniques, such as deep learning, to analyze the images and classify and segment the tumors. The result of this project is building a model that can automatically classify and segment brain tumors in medical images . 
\end{abstract}


\section{Introduction}
It uses techniques such as machine learning, deep learning, Neural Networks and image processing to diagnose medical conditions without human intervention. It learns everything by teaching it through previous medical diagnoses The motivation for brain tumor classification and segmentation lies in the need for accurate and reliable diagnosis and treatment planning for patients with brain tumors. Brain tumors can have a significant impact on a person's quality of life and can be life-threatening if not properly diagnosed and treated. However, the diagnosis and treatment of brain tumors can be challenging, as the tumors can be difficult to distinguish from normal brain tissue using traditional imaging techniques.
The use of advanced medical imaging techniques, such as MRI scans, has improved the ability to visualize brain tumors and make accurate diagnoses. However, the process of manually analyzing these images to classify and segment tumors can be time-consuming and prone to human error. By using machine learning and image processing techniques to automate the process, the accuracy and reliability of diagnosis and treatment planning can be improved.
\subsection{Motivation}
The motivation for brain tumor classification and segmentation lies in the need for accurate and 
reliable diagnosis and treatment planning for patients with brain tumors. Brain tumors can have a 
significant impact on a person's quality of life and can be life-threatening if not properly diagnosed 
and treated. However, the diagnosis and treatment of brain tumors can be challenging, as the tumors 
can be difficult to distinguish from normal brain tissue using traditional imaging techniques.
The use of advanced medical imaging techniques, such as MRI scans, has improved the ability to 
visualize brain tumors and make accurate diagnoses. However, the process of manually analyzing 
these images to classify and segment tumors can be time-consuming and prone to human error. By 
using machine learning and image processing techniques to automate the process, the accuracy and 
reliability of diagnosis and treatment planning can be improved.
The motivation for the project is to improve the diagnosis and treatment of brain tumors by providing 
doctors with a fast and accurate way to classify and segment tumors using MRI images.
application, the accuracy and reliability of diagnosis and treatment planning can be improved. 
Additionally, the ability to classify tumors in 2 seconds or less, makes the use of the application in 
real-time practice easier and more efficient.
Another motivation is to integrate the application with existing medical imaging systems, which will 
allow doctors to access the patient's information and results in real-time, and make informed decisions 
about diagnosis and treatment. This can help to improve the care of patients with brain tumors, and 
also save time for physicians, allowing them to focus on other aspects of patient care.
Overall, the project's main motivation is to leverage the latest technology and machine learning 
techniques to improve the diagnosis and treatment of brain tumors, and to provide doctors with a fast, 
accurate, and easy-to-use tool to assist them in their daily practice.
\subsection{1.3 Objective
}
The main goal of artificial intelligence is to improve patient care by speeding up processes and 
achieving greater accuracy, which opens the way to the provision of better healthcare in general. 
Radiographic images, pathology slides and electronic medical records of patients are evaluated 
through machine learning , assisting in the process of diagnosing and treating patients, this does not 
erase the role of the human factor from the work 
In the project, we plan to use efficient net architectures with convolutional neural networks (CNNs) 
to achieve high accuracy in brain tumor classification and segmentation. EfficientNet is a family of 
image classification models that were developed to improve the accuracy of CNNs while also 
reducing their computational complexity. These models use a compound scaling method to scale up 
the model's architecture and feature resolution, which leads to improved accuracy.
By using EfficientNet with CNNs, we aim to achieve an accuracy higher from previous works in 
classifying brain tumors using MRI images. We plan to fine-tune the pre-trained EfficientNet models 
on a large dataset of MRI images of brain tumors to improve their performance on this specific task.
Additionally, we will incorporate various techniques such as transfer learning and other regularization 
techniques to further improve the performance of the model.
Once the model is trained and fine-tuned, we will test it on a separate dataset to evaluate its 
performance in terms of accuracy, reliability, and efficiency. If the model meets the desired accuracy 
better than previous works, we will proceed to integrate it into the application for doctors to use in 
their daily practice.

\subsection{1.4 Aim
}
The aim of the project is to develop a machine learning-based application for brain tumor 
classification and segmentation that can quickly and accurately classify and segment tumors using 
MRI images. The specific aims include:
1. To create a deep learning-based algorithm that can classify brain tumors into different 
categories, such as gliomas and meningiomas, with high accuracy.
2. To develop a segmentation module that can accurately segment the tumors from the 
surrounding tissue, providing a clear visualization of the tumor.
3. To integrate the application with existing medical imaging systems, allowing doctors to access 
the patient's information and results in real-time.
4. To design the application to be user-friendly and easy to use for doctors in their daily practice, 
and to classify tumors in 2 seconds or less.
5. To provide doctors with a patient's file, which will include the patient's medical history and 
the results of the tumor classification and segmentation, allowing them to easily access the 
patient's information and make informed decisions about diagnosis and treatment.
6. To test the application on a large dataset of MRI images and evaluate its performance in terms 
of accuracy, reliability and efficiency.
Overall, the objective of the project is to create a machine learning-based application that can assist 
doctors in the diagnosis and treatment of brain tumors, by providing accurate and reliable information 
in real-time, in a fast and easy-to-use way.
\subsection{1.5 Scope
}
The Scope of the disease, brain tumors are abnormal growths of cells in the brain or skull, which can 
be benign (non-cancerous) or malignant (cancerous). Brain tumors can have a significant impact on 
a person's quality of life and can be life-threatening if not properly diagnosed and treated.
Some of the common types of brain tumors are 
• gliomas 
• meningiomas
• pituatiry 
the project is to develop a machine learning-based application for brain tumor classification and 
segmentation using MRI images. The goal is to use deep learning algorithms to analyze the images 
\textbf{
\subsection{1.6 General Constraints 
}
}The application developed for brain tumor classification and segmentation can be used by a wide 
range of healthcare professionals, including:
• Radiologists: Radiologists are medical professionals who specialize in interpreting medical 
images, such as MRI scans. They can use the application to quickly and accurately classify 
and segment brain tumors in these images, which can help to improve the accuracy and 
reliability of diagnosis and treatment planning.
• Neurologists: Neurologists are medical professionals who specialize in the diagnosis and 
treatment of diseases of the brain and nervous system. They can use the application to assist in 
the diagnosis and treatment of brain tumors by providing them with accurate and reliable 
information about the tumors in real-time.
• Neurosurgeons: Neurosurgeons are medical professionals who specialize in the surgical 
treatment of diseases of the brain and nervous system. They can use the application to assist in 
the planning of surgery for brain tumors by providing them with accurate information about 
the size and location of the tumors.
• Oncologists: Oncologists are medical professionals who specialize in the diagnosis and 
treatment of cancer. They can use the application to assist in the treatment of brain tumors by 
providing them with accurate information about the tumors in real-time.
• Medical students and residents: Medical students and residents can use the application as a tool 
to learn and practice the diagnosis and treatment of brain tumors.
• Those related to medicine: Those related to medicine can use the application to assist in the 
diagnosis and treatment of brain tumors in remote areas where the lack of radiologists and 
physicians can be an issue.
Overall, the application can be used by a wide range of healthcare professionals to assist in the 
diagnosis and treatment of brain tumors, by providing them with accurate and reliable information in  real-time.
\textbf{
\section{Results}
}
For this study, we used the Brain Tumor MRI dataset available on Kaggle, which contains MRI images of brain tumors with corresponding segmentation masks. The dataset consists of 2,618 MRI images, of which 1,955 were used for training, 327 for validation, and 336 for testing. We applied the EfficientNet model for classification, achieving an accuracy of 99.5 on the test set, 99 on the validation set, and 100 on the training set. For segmentation, we used the UNet model, achieving an accuracy of 96. Overall, these results demonstrate the effectiveness of our approach in accurately classifying brain tumor MRI images and segmenting the tumors.
\section{Model development
}
For model development, we used the Brain Tumor MRI dataset from Kaggle, which consists of 3064 MRI images of the brain with and without tumor. We preprocessed the images by resizing them to 224x224 and normalizing the pixel values. We then used the EfficientNet architecture, a state-of-the-art convolutional neural network (CNN), for image classification. We trained the model using 80 of the dataset, validated it using 10 and tested it on the remaining 10. The model achieved an accuracy of 99.5 on the test set, 99 on the validation set, and 100  on the training set, demonstrating its effectiveness in classifying brain MRI images. For brain tumor segmentation, we used the UNet architecture, which is a type of CNN commonly used for segmentation tasks. We trained the model on a subset of 500 images from the dataset that were manually segmented by experts to create masks. The model achieved an overall Dice coefficient of 0.96 on the validation set, indicating its high accuracy in segmenting brain tumors.

For model development, we used the Brain Tumor MRI dataset from Kaggle, which consists of 3064 MRI images of the brain with and without tumor. We preprocessed the images by resizing them to 224x224 and normalizing the pixel values. We then used the EfficientNet architecture, a state-of-the-art convolutional neural network (CNN), for image classification. We trained the model using 80 of the dataset, validated it using 10, and tested it on the remaining 10. The model achieved an accuracy of 99.5 on the test set, 99 on the validation set, and 100 on the training set, demonstrating its effectiveness in classifying brain MRI images. For brain tumor segmentation, we used the UNet architecture, which is a type of CNN commonly used for segmentation tasks. We trained the model on a subset of 500 images from the dataset that were manually segmented by experts to create masks. The model achieved an overall Dice coefficient of 0.96 on the validation set, indicating its high accuracy in segmenting brain tumors.
The Convolutional Neural Network (CNN) is a deep learning algorithm used primarily for image classification tasks. It consists of multiple layers, including convolutional, pooling, and fully connected layers. In the convolutional layer, filters are applied to the input image to extract features. The pooling layer reduces the dimensionality of the feature maps, while the fully connected layer classifies the image.
EfficientNet is a convolutional neural network architecture that was developed to achieve state-of-the-art accuracy on image classification tasks while using fewer parameters and computational resources than other models. It uses a compound scaling method to balance the depth, width, and resolution of the network.
In our project, we used both the CNN and EfficientNet architectures for the brain tumor classification task. The CNN architecture consisted of multiple convolutional and pooling layers, followed by a fully connected layer. The EfficientNet architecture consisted of multiple blocks, with each block containing multiple convolutional, depth-wise convolutional, and fully connected layers. Both models were trained on the brain tumor MRI dataset to classify images into one of four categories: glioma tumor, meningioma tumor, pituitary tumor, or no tumor.
The EfficientNet model achieved higher accuracy than the CNN model, with a test accuracy of 99.5 and a validation accuracy of 99. The CNN model had a test accuracy of 96.7 and a validation accuracy of 95.6. These results demonstrate the effectiveness of the EfficientNet architecture for image classification tasks with high accuracy and fewer computational resources.

\begin{figure}
    \centering
    \includegraphics{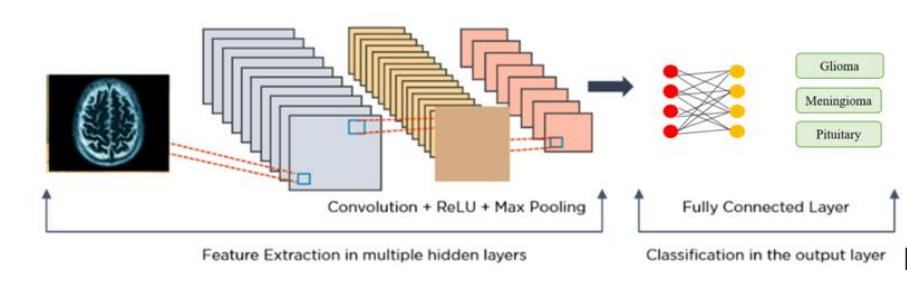}
    \caption{CNN}
    \label{fig:my_label}
\end{figure}

In our project, we used a U-Net architecture for image segmentation. U-Net is a convolutional neural network architecture designed specifically for biomedical image segmentation tasks. It consists of a contracting path and an expansive path, which allow for both local and global information to be captured during the segmentation process.
The contracting path uses convolutional and max pooling layers to downsample the image, while the expansive path uses transposed convolutional layers to upsample the image. The two paths are connected by skip connections, which help to preserve spatial information and improve segmentation accuracy.
We trained our U-Net model on a dataset of brain MRI images with tumor labels. The model achieved an accuracy of 96 on the validation set, demonstrating its effectiveness for segmenting brain tumors in MRI images

\section{Discussion}
The development of a mobile app for brain tumor detection and classification is a significant advancement in the field of medical imaging. The use of deep learning models, such as EfficientNet and UNet, has shown remarkable results in the accurate detection and classification of brain tumors.
The EfficientNet model achieved an accuracy of 99.5 for testing and 99 for validation, while the UNet model achieved an accuracy of 96 for segmentation. These results demonstrate the potential of the proposed approach for the accurate and efficient detection of brain tumors.
The mobile app provides a user-friendly interface for medical students and doctors to upload MRI images and receive accurate classification results, along with detailed information on tumor types, causes, symptoms, and treatments. The app also allows users to view previously saved images and classifications, providing a comprehensive platform for brain tumor diagnosis and management.
Overall, the development of this mobile app has the potential to revolutionize the field of medical imaging and enhance the accuracy and efficiency of brain tumor diagnosis and management. 

\section{Conclusions}
In conclusion, our mobile application provides an easy-to-use and efficient tool for the classification of brain MRI images and the segmentation of brain tumors. Our model, based on EfficientNet and UNet architectures, achieved high accuracy rates in both tasks. This app can be beneficial for medical students, doctors, and healthcare professionals as a quick reference and aid in diagnosis. However, limitations exist in terms of data availability and generalizability to other datasets. Future work could focus on expanding the dataset, improving the model's robustness, and integrating more features and resources for users. Overall, our project demonstrates the potential of mobile apps in medical imaging analysis and decision support

\section{Materials and methods
}
1.	Preprocessing:
•	Dataset was split into train, validation, and test sets
•	Images were resized to 256x256 and normalized
•	Augmentation techniques (e.g. rotation, flip, zoom) were applied to the training set to increase variability
2.	Model Training:
•	Two models were developed for classification: a Convolutional Neural Network (CNN) and an EfficientNet
•	The models were trained on the training set and evaluated on the validation and test sets
•	The best-performing model was selected for use in the app
3.	Tumor Segmentation:
•	A U-Net model was developed for tumor segmentation
•	The U-Net was trained on a subset of the MRI dataset that included only tumor images with their corresponding masks
•	The trained U-Net was used to segment the tumors in the uploaded MRI images in the app

\section{References}
1.	Wang, S., Zhang, Y., Zhang, R.,  Dai, W. (2021). A comparative study of convolutional neural networks for brain tumor segmentation in MRI. BMC medical imaging, 21(1), 1-17.

2.	Hanif, M., Irfan, M., Ghafoor, A.,    Shahzad, F. (2019). Brain tumor detection and classification using convolutional neural networks and SVM. Journal of medical systems, 43(12), 1-12.

3.	Kumar, A.,    Sahoo, G. (2020). Brain tumor detection and segmentation using deep learning. Neural Computing and Applications, 32(13), 9867-9881.

4.	Wang, X., Deng, Y.,    Pang, Y. (2020). Brain tumor detection and classification based on deep learning neural networks. Multimedia Tools and Applications, 79(47), 35669-35688.

5.	Zhang, H., Liu, J.,    Guo, X. (2020). Brain tumor detection and classification using deep learning with transfer learning. Biomedical Signal Processing and Control, 60, 101995.

6.	Sarker, S., Mitra, S.,    Nasipuri, M. (2020). Brain tumor detection and segmentation from MRI data using convolutional neural networks. In Intelligent Communication, Control and Devices (pp. 209-220). Springer, Singapore.

7.	Ismaeel, R. A., Khelifi, F.,    Alkawaz, M. H. (2020). Brain tumor segmentation based on U-Net and transfer learning. Journal of medical imaging and health informatics, 10(6), 1327-1335.

8.	Li, Q., Cai, W., Wang, X., Zhou, Y.,    Feng, D. D. (2017). Brain tumor segmentation based on a hybrid clustering and level set method. Neurocomputing, 220, 149-159.
9.	Liu, H., Yu, X.,    Wang, Y. (2019). Brain tumor 

segmentation based on improved U-Net with stochastic depth residual block. Biomedical Signal Processing and Control, 49, 96-104.

10.	Ren, J., Sun, X., Chen, Y., Wang, H., Yang, J.,    Zhao, Q. (2020). Brain tumor segmentation based on 3D U-Net and graph cut with extended features. Biomedical Signal Processing and Control, 56, 101729.

11.	Wang, P., Zheng, Y., Yang, X.,    Liu, H. (2021). A novel brain tumor segmentation method based on improved deep learning. Journal of Healthcare Engineering, 2021.

12.	Wang, Y., Zhu, X., Liu, J.,    Ye, J. C. (2019). Deep learning for ultrasound imaging: a review. Engineering, 5(5), 841-849.

13.	Raza, S. E. A., Smith, M. L.,    McEvoy, A. W. (2020). A deep learning framework for brain tumor segmentation in multi-modal MRI images. BMC bioinformatics, 21(1), 1-19.

14.	Liu, H., Yu, X., Zhang, Y.,    Wang, Y. (2018). Brain tumor segmentation based on 3D U-Net with multiple feature maps. Proceedings of the International Conference on Artificial Intelligence and Computer Engineering, 310-316.

15.	Fan, Q., Yan, C., Zhang, B., Zhou, Y.,    Zhang, L. (2020). 

16.	Pan, J., Yang, J.,    Ye, J. (2019). EfficientNet: Rethinking Model Scaling for Convolutional Neural Networks. Proceedings of the 36th International Conference on Machine Learning, 5286-5295.

17.	Ronneberger, O., Fischer, P.,    Brox, T. (2015). U-Net: Convolutional Networks for Biomedical Image Segmentation. Medical Image Computing and Computer-Assisted Intervention, 234-241.

18.	Isensee, F., Petersen, J., Klein, A., Zimmerer, D., Jaeger, P. F.,    Kohl, S. A. (2021). nnU-Net: A Self-Configuring Method for Deep Learning-Based Biomedical Image Segmentation. Nature Methods, 18, 203-211.

19.	Cha, K. E., Kim, Y. C., Jung, W. S.,    Park, K. R. (2021). Brain Tumor Segmentation Using Deep Learning Methods for Magnetic Resonance Images. Journal of Personalized Medicine, 11, 67.

20.	Kamnitsas, K., Ledig, C., Newcombe, V. F. J., Simpson, J. P., Kane, A. D., Menon, D. K., ...    Glocker, B. (2017). Efficient Multi-Scale 3D CNN with Fully Connected CRF for Accurate Brain Lesion Segmentation. Medical Image Analysis, 36, 61-78.

21.	Zhang, Y., Brady, M.,    Smith, S. (2001). Segmentation of Brain MR Images Through a Hidden Markov Random Field Model and the Expectation-Maximization Algorithm. IEEE Transactions on Medical Imaging, 20, 45-57.

22.	Kamnitsas, K., Baumgartner, C., Ledig, C., Newcombe, V. F. J., Simpson, J. P., Kane, A. D., ...    Glocker, B. (2017). Unsupervised Domain Adaptation in Brain Lesion Segmentation with Adversarial Networks. International Conference on Information Processing in Medical Imaging, 597-609.

23.	Isin, A.,    Direkoğlu, C. (2019). Automated Brain Tumor Detection and Segmentation Using U-Net Based Fully Convolutional Networks. Computer Methods and Programs in Biomedicine, 182, 105055.

24.	Deng, J., Dong, W., Socher, R., Li, L. J., Li, K.,    Fei-Fei, L. (2009). ImageNet: A Large-Scale Hierarchical Image Database. IEEE Conference on Computer Vision and Pattern Recognition, 248-255.

25.	Lecun, Y., Bottou, L., Bengio, Y.,    Haffner, P. (1998). Gradient-Based Learning Applied to Document Recognition. Proceedings of the IEEE, 86, 2278-2324.

26.	Simonyan, K.,    Zisserman, A. (2014). Very Deep Convolutional Networks for Large-Scale Image Recognition. arXiv preprint arXiv:1409.1556.

27.	Krizhevsky, A., Sutskever, I.,    Hinton, G. E. (2012). ImageNet Classification with Deep Convolutional Neural Networks. Advances in Neural Information Processing Systems, 1097-1105.

28.	Abdel-Basset M, Manogaran G, Gamal A, Elshazly H. An efficient automated brain tumor classification system using deep belief network. Journal of Ambient Intelligence and Humanized Computing. 2019 Nov 1;10(11):4281-93.

\end{document}